\documentclass[epj]{svjour}
\usepackage{graphicx}
\begin{document}
\title{The role of second viscosity in the velocity shear-induced heating mechanism}
\author{G. Tsereteli \and Z. Osmanov% etc
}                     % Do not remove
\offprints{}          % Insert a name or remove this line
\institute{School of Physics, Free University of Tbilisi, 0183, Tbilisi,
Georgia}
\date{Received: date / Revised version: date}
\abstract{
in the present paper we study the influence of second viscosity on non-modally induced heating mechanism. For this purpose we study the set of equations governing the hydrodynamic system. In particular, we consider the Navier Stokes equation, the continuity equation and the equation of state, linearise them and analyse in the context of non-modal instabilities. Unlike previous studies in the Navier Stokes equation we include the contribution of compressibility, thus the second viscosity. By analysing several typical cases we show that under certain conditions the second viscosity might significantly change efficiency of the mechanism of heating.
\PACS{
      {PACS-key}{discribing text of that key}   \and
      {PACS-key}{discribing text of that key}
     } % end of PACS codes
} %end of abstract
\maketitle
\section{Introduction}
\label{Theory}

It is well known that in many astrophysical scenarious the flows are characterized by complex kinematics
with inhomogeneous velocity fields, the so called shear flows (SF). The typical examples are the extragalactic jets \cite{broder,kharb} and the young stellar jets \cite{yso}, which reveal tornado-like structures of kinematics. Such a complexity is observationally supported also in solar spicules \cite{pm98}. Therefore, it is clear that under certain conditions the SFs  might somehow influence the plasma processes in the mentioned astrophysical situations.

It has been realised that even in relatively simple velocity fields the SFs lead to instabilities \cite{modal}.
In the framework of the standard modal approach the physical system is governed by equations having the form $\partial_t\xi+\left({\bf V_0\cdot\nabla}\right)\xi = M \xi$. Here, $M$ denotes a Hermitian differential operator, ${\bf V_0}$ is the background velocity and $\xi$ is the field from the Hilbert space. In this scheme a hydrodynamical picture is decomposed into a system of independent normal modes with certain time constants. For example, the linearized set of equations can be solved by using an anzatz, with a well known time dependence $\sim e^{-i\omega t}$, where $\omega$ denotes the frequency of the corresponding oscillations. In the framework of this approach all physical quantities nontrivially depend on spacial coordinates and the SFs lead to a non-Hermitian operator, $M$ \cite{tatsuno}. This means that the non-modally excited waves are more complex and might be very interesting, which is the main reason why we examine this particular class of instabilities.

The corresponding mathematical model for non-modal waves has been developed by Lord Kelvin \cite{kelvin}. In the framework of this approach, if one uses a specially chosen anzatz, the physical quantities expanded in the linear approximation satisfy equations which are reduced to the ordinary differential equations with initial value problem \cite{tref}.

In \cite{chven} we have studied
the role of velocity shear induced phenomena in magnetohydrodynamic flows. It has been shown that by means of the non-modal
instability the physical system exhibits the energy exchange between slow/fast magnetosonic and
Alfv\'en waves. Electrostatic ion perturbations has been considered in \cite{electrost}. Authors have
found that depending on shear parameters, the system undergoes two different kinds of instabilities. For the exponentially evolving
wave vectors the electrostatic perturbations evolve exponentially as well. On the other hand, when wave vectors are limited in amplitudes
and the flow is mostly stable, for a certain set of parameters the physical system might exhibit the so called
parametric instability, which takes place only for relatively narrow ranges of physical parameters.

In this paper we study another interesting consequence of the shear-induced instability. This problem originally has been examined in \cite{andro} where the author considers the process of heating as to be composed of three major steps: 1. excitation of waves; 2. their non-modal amplification and 3. dissipation in the form of heat by means of the viscosity. The role of viscosity is crucial, although non trivial. In particular, it is clear that dissipation is more efficient for more viscous fluids, but on the other hand, in the medium with high viscosity the waves cannot amplify significantly extracting energy from the means flow. Therefore, only for relatively moderate values of viscosity the non-modal self-heating might be efficient.

A direct application of this mechanism to astrophysical problems has been presented in \cite{rop10}, where the problem was studied in the context of the well known chromospheric heating of solar-type stars. It has been shown that acoustic waves might undergo the non-modal SF instability, leading to damping via the viscous dissipation. We have found that this mechanism might explain  unusually high temperature of chromosphere of slowly rotating stars. The similar problem, but for MHD flows has been studied in \cite{orp12}, where together with viscosity the magnetic resistivity was included. We have shown that the velocity shear complexity still might lead to the non-modal amplification of waves, finally leading to efficient heating of the background flow.

In the aforementioned articles the self-heating was studied for incompressible fluids and therefore the role of the second viscosity (which is crucial for compressible fluids) has not been studied. On the other hand, it is natural to generalise the previous study by including in the Navier Stokes equation the terms corresponding to compressibility and consider efficiency of self-heating.

The paper is organised in the following way. In section~II, a theoretical model of shear-induced self-heating mechanism is presented, in section~III we discuss the obtained numerical results and in section ~IV we summarise them.

\section{Main consideration}
To conduct a detailed analysis of the viscous self-heating of non-modally generated waves and study the influence of second viscosity on the system's behavior, we consider the standard set of hydrodynamic equations.
The mass conservation:
\begin{equation}
D_t\rho+\rho {\bf \nabla} \cdot {\bf V}=0,
\end{equation}
the momentum conservation, taking into account the second viscosity
\begin{equation}
D_t {\bf V}=-1/\rho {\bf \nabla} P+\eta\Delta {\bf V}+ (\eta/3+\zeta){\bf \nabla}({\bf \nabla} \cdot {\bf V})
\end{equation}
and the polytropic equation of state
\begin{equation}
P=C\rho ^n,
\end{equation}
where $D_t\equiv\partial_t+({\bf V} \cdot {\bf \nabla})$ is a notation for the convective derivative, $\rho$  is density, $V$ is velocity, $n$ is the polytropic index and $\eta$  and $\zeta$  are respectively coefficients of kinematic shear viscosity and kinematic compression viscosity.
After expanding all physical quantities up to the first order terms
\begin{equation}
\rho\equiv\rho_0+\rho',
\end{equation}
\begin{equation}
{\bf V}\equiv {\bf V_0}+{\bf V'},
\end{equation}
\begin{equation}
P\equiv P_0+P',
\end{equation}
where $\rho_0$, $V_0$ and $P_0$ are the zeroth order quantities and $\rho'$, $V'$ and $P'$ are the corresponding linear terms, which satisfy the following linearized set of equations.
\begin{equation}
\mathcal{D}_t\rho'+\rho_0({\bf \nabla}\cdot {\bf V'})=0,
\end{equation}
\begin{equation}
\mathcal{D}_t{\bf V'}+({\bf V'}\cdot{\bf \nabla}){\bf V_0}=-C_s/\rho_0{\bf \nabla}\rho'+\eta{\bf \Delta V'}+(\eta/3+\zeta)\nabla({\bf \nabla}\cdot {\bf V'}).
\end{equation}
Here $\mathcal{D}_t=\partial_t+({\bf V_0}\cdot {\bf \nabla})$ and $ C_s=\sqrt{dP_0/d\rho_0}$ is the sound speed. Next we follow the approach originally developed in \cite{mahrog} and we expand the velocity by the Taylor series in the vicinity of a point A($x_0$,$y_0$,$z_0$) preserving only the linear terms.
\begin{equation}
V=V(A)+\sum_{i=1}^{3}\frac{\partial V(A)}{\partial x_i}(x_i-x_{i0}),
\end{equation}
In \cite{mahrog} it has been shown that the set of linearized partial differential equations can be transformed into ordinary differential equations by employing the following ansatz
\begin{equation}
F(x,y,z,t)\equiv\hat{\hat{F}}(t)e^{\phi_1-\phi_2}
\end{equation}
\begin{equation}
\phi_1\equiv\sum_{i=1}^3K_i(t)x_i
\end{equation}
\begin{equation}
\phi_2\equiv\sum_{i=1}^3V_i(A)\int K_i(t)dt
\end{equation}
where $V_i(A)$ is the unperturbed velocity component and $K_i(t)$ are the components of wave vectors.

By following the standard approach developed in \cite{mahrog} it is straightforward to show that the convective derivative of a physical quantity $F(x,y,z,t)$
$$\;\;\;\;\;\mathcal{D}_t F=e^{i\left(\phi_1-\phi_2\right)}\partial_t\hat{F}(t)+$$ $$+ix\left(K_x^{(1)}+a_1K_x+b_1K_y+c_1K_z\right)+$$
$$+iy\left(K_y^{(1)}+a_2K_x+b_2K_y+c_2K_z\right)+$$
\begin{equation}\label{conv2}
+iz\left(K_z^{(1)}+a_3K_x+b_3K_y+c_3K_z\right),\end{equation}
becomes a simple time derivative multiplied by $e^{i\left(\phi_1-\phi_2\right)}$ if the following set of equations is satisfied
\begin{equation}\label{dk}
{\bf \partial_{t}K}+ {\bf S^T} \cdot {\bf K}=0,\end{equation}

where
\begin{equation}\label{S}
 {\bf S} = \left(\begin{array}{ccc} V_{x,x} & V_{x,y} & V_{x,z}  \\
V_{y,x} & V_{y,y} & V_{y,z}  \\ V_{z,x} & V_{z,y} & V_{z,z} \\
\end{array} \right )\equiv\left(\begin{array}{ccc} a_1 & a_2 & a_3 \\
b_1 & b_2 & b_3 \\ c_1 & c_2 & c_3 \\
\end{array} \right ),\end{equation}
is the so-called shear matrix and ${\bf S^T}$ is its transposed form. In dimensionless quantities the complete set of equations writes as
\begin{equation}
d^{(1)}+{\bf k}\cdot {\bf u}=0
\end{equation}
\begin{equation}
{\bf u}^{(1)}+{\bf s}\cdot {\bf u}={\bf k}d-\nu_1 k^{2}{\bf u}-\nu_2{\bf k}\left({\bf k}\cdot {\bf u}\right)
\end{equation}
\begin{equation}
{\bf k}^{(1)}+{\bf s}^{T} \cdot {\bf k}=0
\end{equation}
Notations are as follows $d\equiv -i\frac{\rho '}{\rho_0}$, ${\bf u}\equiv \frac{{\bf V'}}{C_s}$, ${\bf k}\equiv \frac{{\bf K}}{K_z(0)}$ , $\nu_{1,2}\equiv \bar{\nu}_{1,2}\frac{K_n}{C_s}$, $\bar{\nu}_{1}\equiv\eta$, $\bar{\nu}_{2}\equiv \eta/3+\zeta$, $\textbf{s}\equiv \textbf{S}/(K_z(0)C_s)$. By $\Psi^{(1)}$ we denote the derivative with respect to dimensionless time $\tau\equiv K_z(0)C_s t$.

\begin{figure}[h]
\resizebox{0.5\textwidth}{!}{%
  \includegraphics{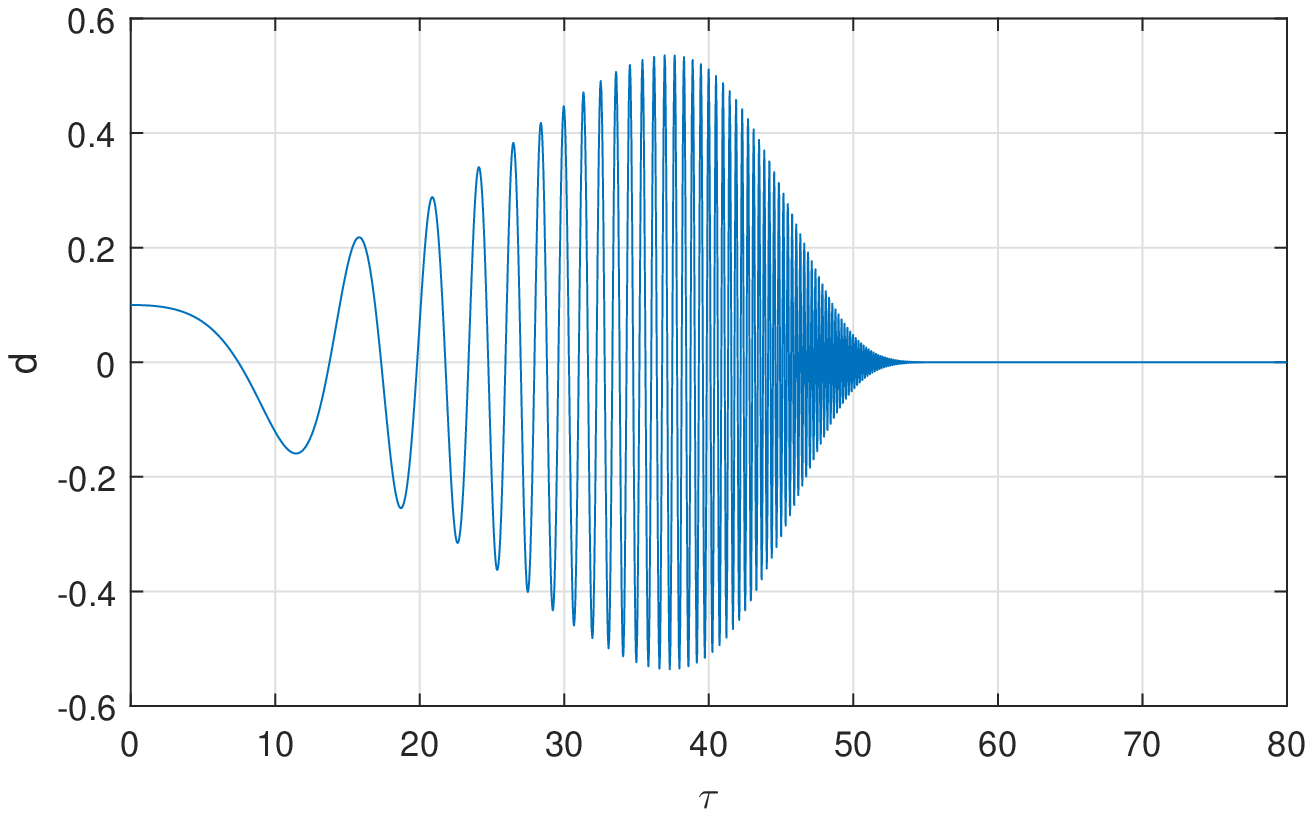}
  \includegraphics{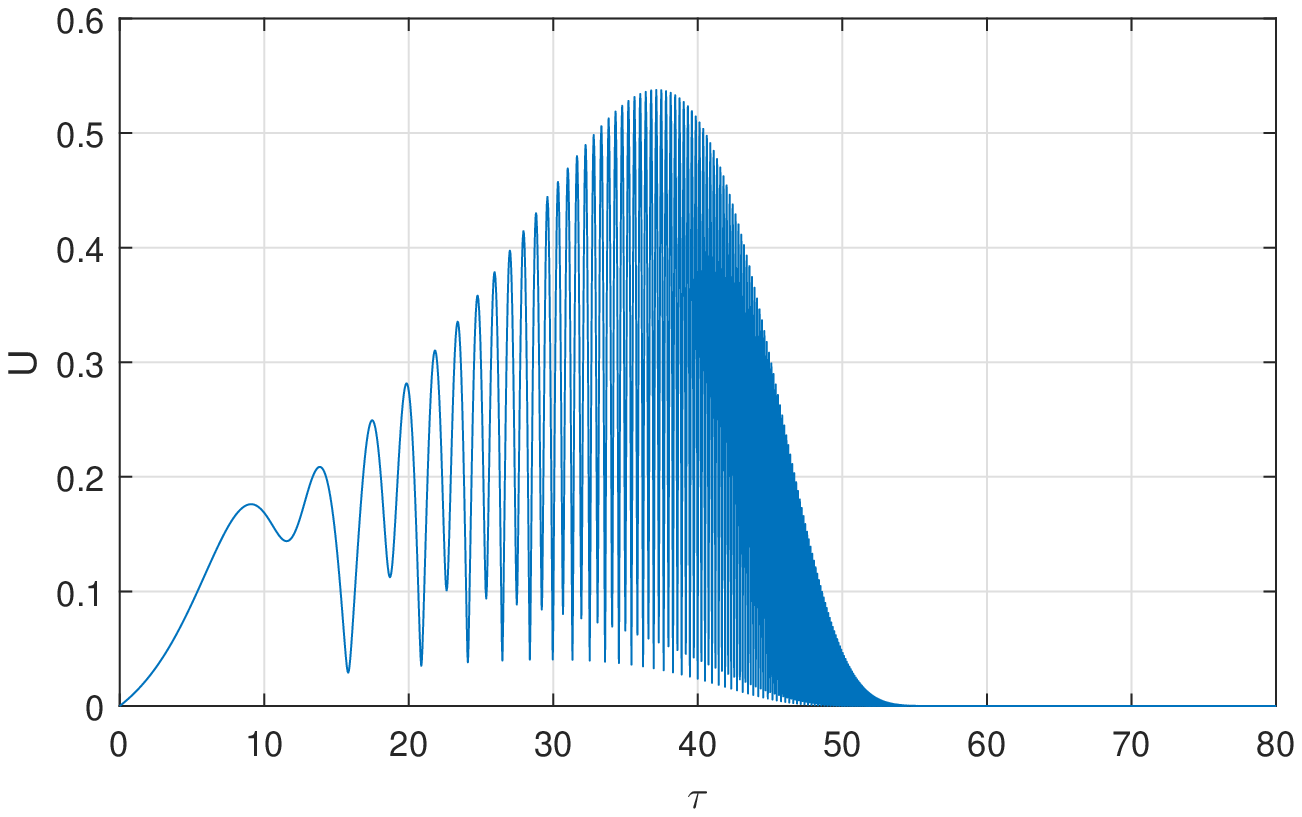}
 }
 \caption{The behavior of $d(\tau)$ and $u(\tau)$ is shown. The set of parameters is $a_1=-0.1$, $a_2 = 0.2$ and the rest of the shear matrix elements is zero, $k_{x}=0.1$, $k_{y}=k_{z}=0$, $u_{x}=u_{y}=u_{z}=0$, $d=0.1$, $\eta=0.001$, $\zeta=0.003$.}
 \label{fig1}
\end{figure}
 \begin{figure}[h]
\resizebox{0.5\textwidth}{!}{%
  \includegraphics{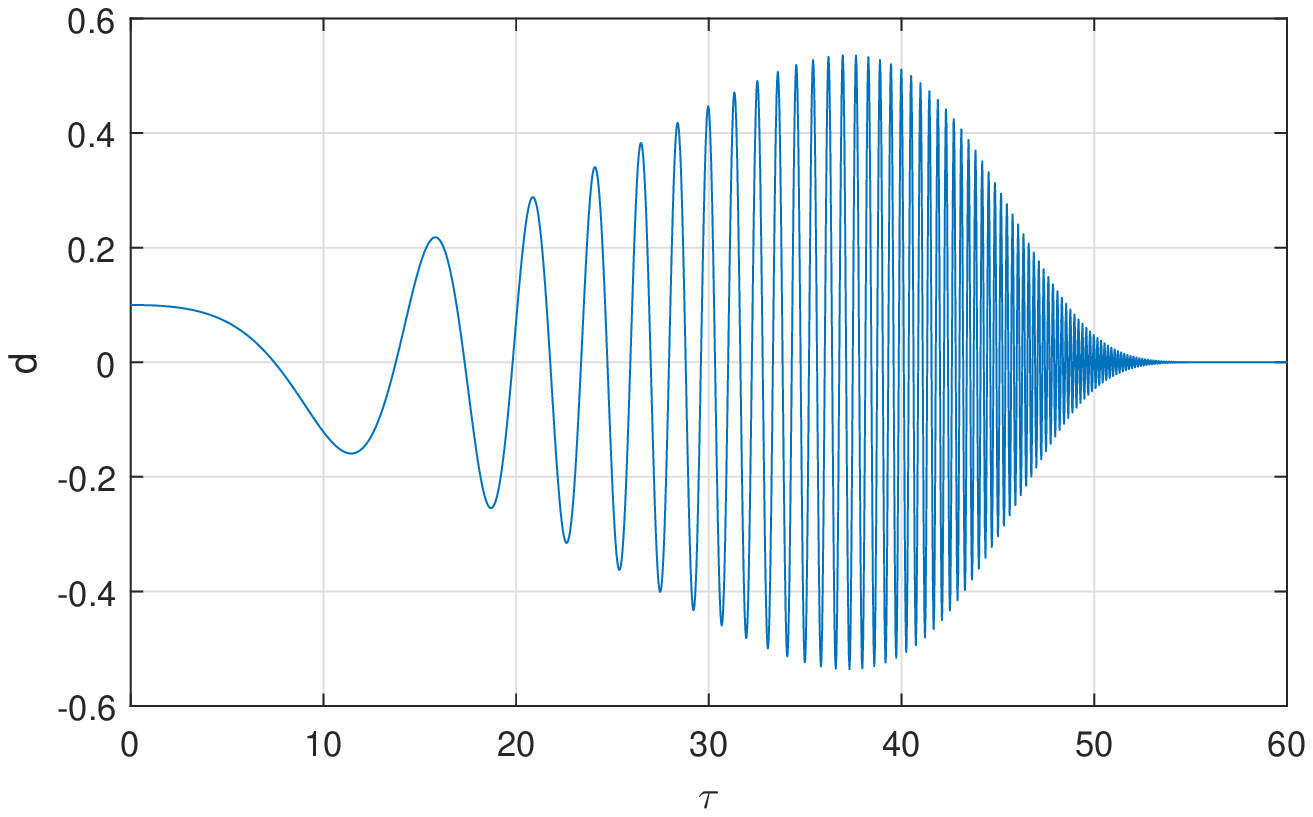}
  \includegraphics{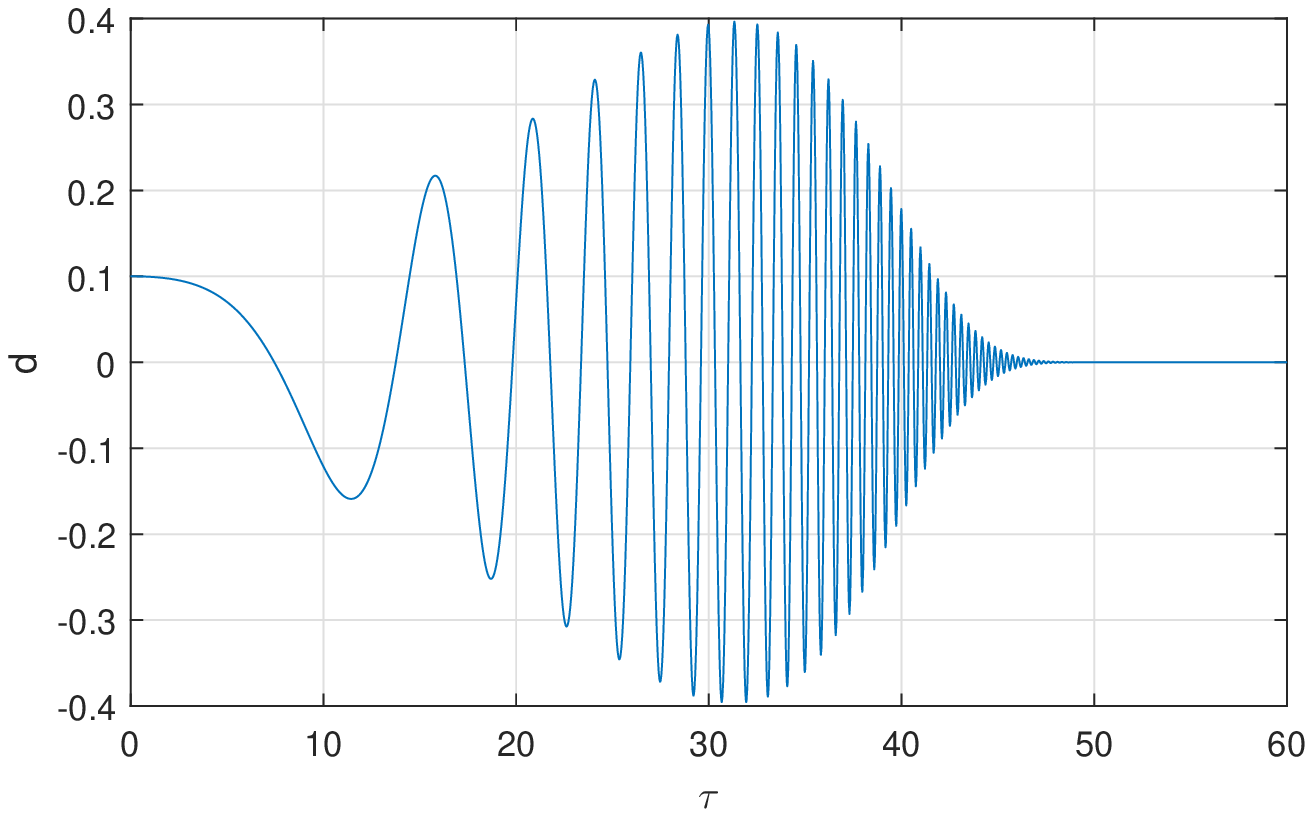}
 }
 \caption{Here we show the plots of density perturbation versus time for the same parameters as in the previous figure, except $\zeta=0$ (left panel) and $\zeta=0.003$ (right panel). }
 \label{fig2}
\end{figure}

In order to analyse the process of heating by means of energy we define the perturbation energy density \cite{landau}
\begin{equation}
 E_{tot}=\frac{{\bf u}^2}{2}+\frac{d^2}{2},
\end{equation}
where the first and second terms are the so-called kinetic and compressional terms. It is straightforward to show that time derivative of $E_{tot}$ is given by
\begin{equation}
 E^{(1)}=-{\bf u}(\textbf{s} \cdot {\bf u})-\nu_1 k^2{\bf u}^2-\nu_2\left({\bf k}\cdot {\bf u}\right)^2.
\end{equation}

As it is evident from the above equation, the last two terms are responsible for heating, therefore efficiency of the process might be defined as follows \cite{andro}
$$\psi(\tau)=$$
\begin{equation}
=\frac{1}{E_{tot}(0)}\int^{\tau}_{0}\left(\nu_1k^2(\tau') u^2(\tau')+\nu_2\left({\bf k}(\tau')\cdot {\bf u}(\tau')\right)^2\right)d\tau',
\end{equation}
where $E_{tot}(0)$ is the initial energy of perturbation.
\section{Discussion}
 \begin{figure}[h]
\resizebox{0.5\textwidth}{!}{%
  \includegraphics{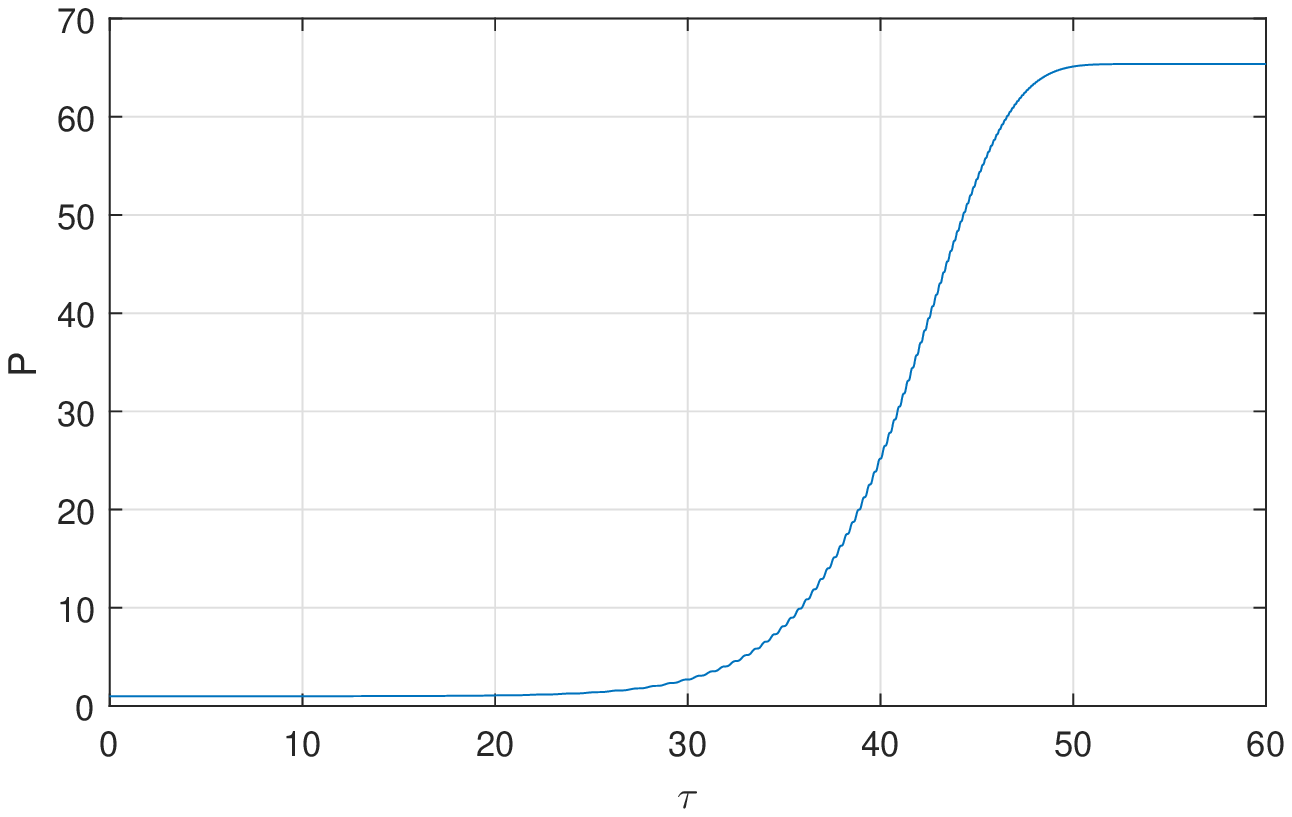}
  \includegraphics{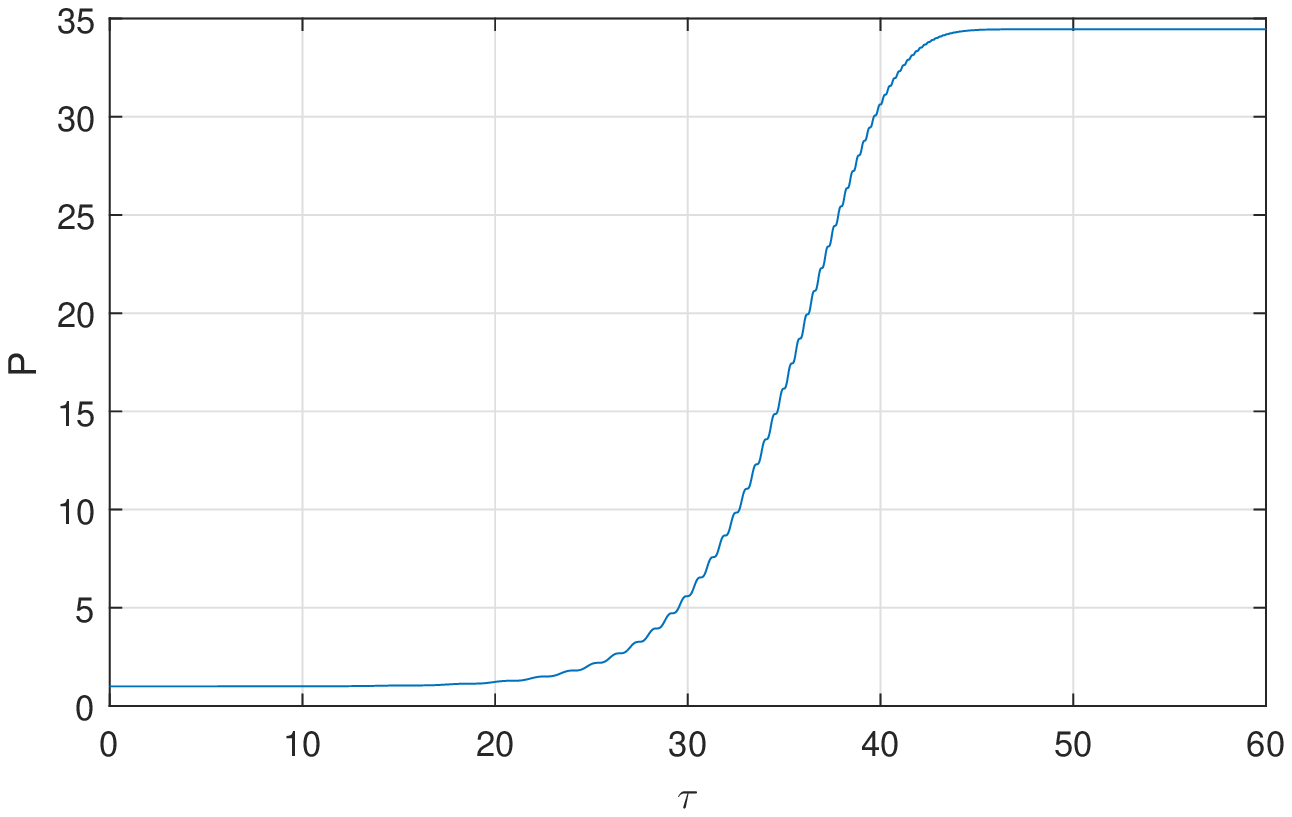}
 }
 \caption{Here we demonstrate efficiency of heating versus time.  The set of parameters is the same as in the previous figure: left panel corresponds to $\zeta=0$ and the right panel - to $\zeta=0.003$.}
 \label{fig3}
\end{figure}

In this section we are going to numerically demonstrate the effect of compression on the overall heating for arbitrary parameters in reasonable scopes. We start with the set of equation for different types of SFs. For the incompressible case the shear matrix is relatively trivial and is traceless. In particular, from Eq. 1 it is evident that from $\rho = const$ follows the condition ${\bf \nabla}\cdot {\bf V}=0$, which if rewritten for the matrix components means $a_1+b_1+c_1 = 0$. In all of the briefly discussed articles in the introduction the problem of heating has been examined for the incompressible fluids. One can straightforwardly show that the same condition for the compressible flows is not satisfied at all and therefore, it is important to consider this particular case as well.

\subsection{Density perturbation}
As a first example we consider a case when initially only density field is perturbed, thus an acoustic wave is excited. In Fig. 1 we demonstrate the behavior of $d(\tau)$ and $u(\tau)$. The set of parameters is $a_1=-0.1$, $a_2 = 0.2$ and the rest of the shear matrix elements is zero, $k_{x}=0.1$, $k_{y}=k_{z}=0$, $u_{x}=u_{y}=u_{z}=0$, $d=0.1$, $\eta=0.001$, $\zeta=0.003$. Without going into details we consider the case $\zeta/\eta = 3$, which is a reasonable value for gasses \cite{cramer}.

As it is clear from the plots, acoustic waves excite velocity components and amplitude of corresponding oscillations initially increase but in due course of time the induced instability is reduced and finally damped by means of the viscosities.

In Fig. 2 we show how sensitive behavior of excited waves might be to the change of second viscosity. In particular, if the instability is almost completely damped at $\tau\approx 55$ for $\zeta=0$ (right panel), the corresponding oscillations for $\zeta=0.003$ (left panel) dissipate at $\tau\approx 45$. Therefore, the second viscosity has a visible effect on the propagation of the acoustic waves and will inevitably affect the nature of the flow. In Fig. 3 we show the time dependence of $\psi$ on time. It is clear that the heating rate is a continuously increasing function of time with relatively high initial increment. This is a direct consequence of the fact that initially the waves are efficiently amplifying, which in turn results in the efficient energy transform to heat. In due course of time, by means of viscous effects the oscillations start damping and consequently a value of $\psi$ saturates. As it is clear from the plots, for the mentioned physical parameters, efficiency of heating is almost two times smaller for $\zeta = 0.003$ than for $\zeta = 0s$.

\subsection{Velocity perturbation}
  \begin{figure}[h]
\resizebox{0.5\textwidth}{!}{%
  \includegraphics{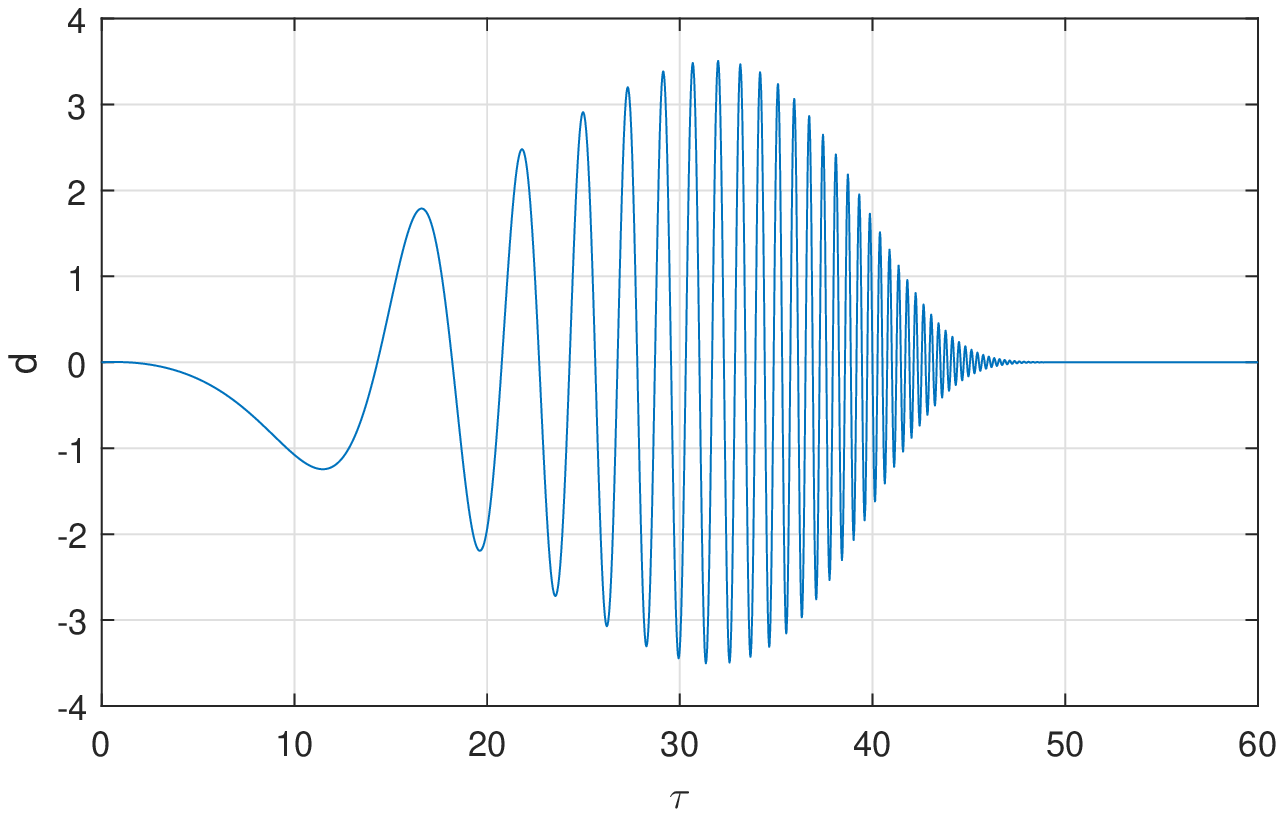}
  \includegraphics{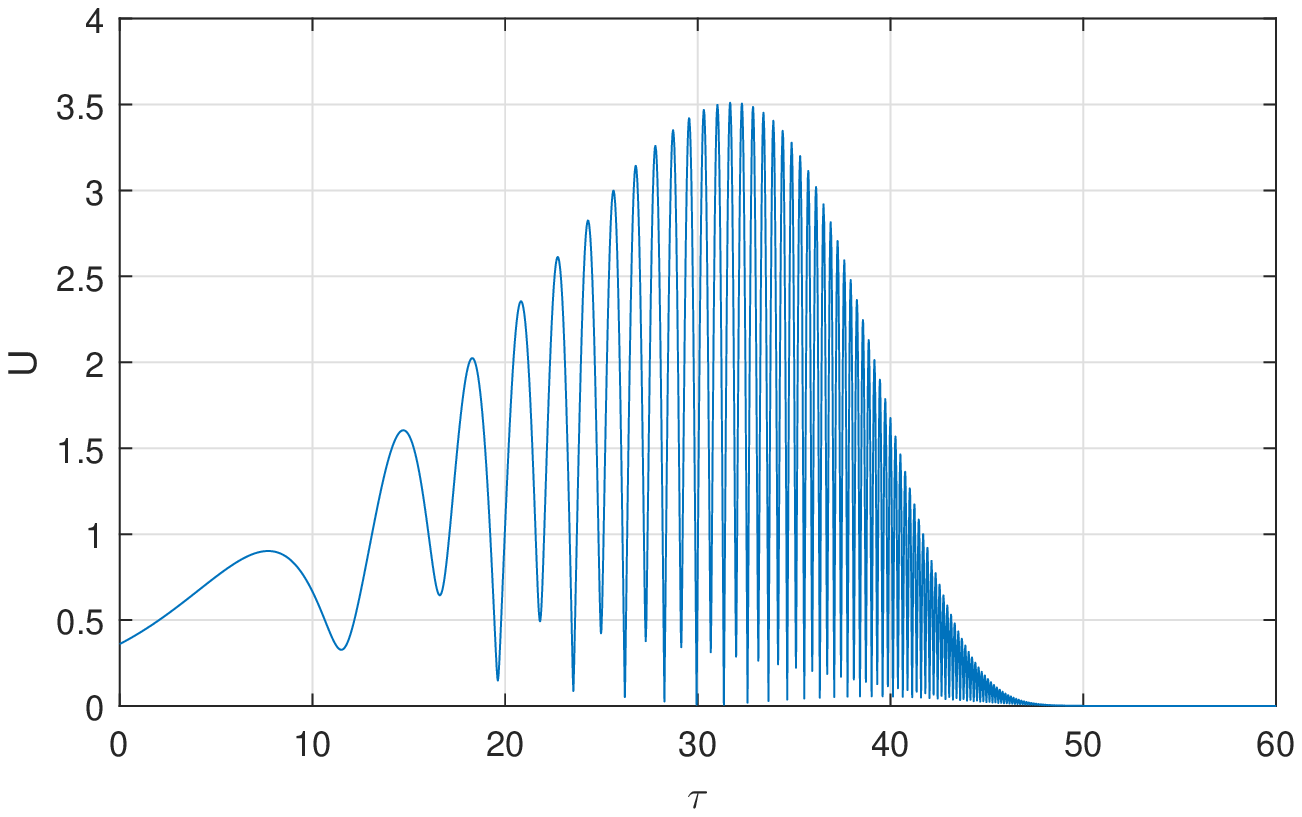}
 }
 \caption{The plots of $d(\tau)$ and $u(\tau)$ are shown. The set of parameters is $a_1=-0.1$, $a_2 = 0.2$ and the rest of the shear matrix elements is zero and $k_{x}=0.1$, $k_{y}= 0.2$, $k_{z}=0$, $u_{x}=0.3$, $u_{y}=-0.2$, $u_{z}=0$, $d=0$, $\eta=0.001$, $\zeta=0.003$.}
 \label{fig4}
\end{figure}

In Fig. 4 we show the plots of $d(\tau)$ and $u(\tau)$, but unlike the Fig. 1, there is no initial density perturbation, but instead the velocity components are excited. The set of parameters is $a_1=-0.1$, $a_2 = 0.2$ and the rest of the shear matrix elements is zero and $k_{x}=0.1$, $k_{y}= 0.2$, $k_{z}=0$, $u_{x}=0.3$, $u_{y}=-0.2$, $u_{z}=0$, $d=0$, $\eta=0.001$, $\zeta=0.003$. It is clear that like the previous results, when the initially perturbed density excites the velocity field, in the current plots the latter excites the acoustic waves. Finally, as previously, by means of the both viscosity coefficients the unstable waves completely dissipate, which means that the flow will be heated.

In Fig. 5  we demonstrate rate of heating for two different values of $\zeta$: $0$ (left panel) and $0.003$ (right panel). Other parameters are the same as in Fig. 4. It is clear that by neglecting the second viscosity the corresponding heating rate is significantly different from a realistic value.

 \section{Conclusion}
We have considered the compressible fluids and studied the role of second viscosity in the process of heating. In particular, we have examined the Navier Stokes equation, continuity equation and equation of state, linearised them and investigated the generation of non-modal SF instabilities.

It has been shown that initially generated acoustic waves perturb velocity components as well and as a result non-modal waves amplify extracting energy from the mean flow. The already amplified waves continuously damp by means of viscosity.

For relatively simple examples we have shown that heating rate might be significantly changed by means of the second viscosity coefficient. In particular, it has been shown that initially driven acoustic waves finally dissipate and the corresponding efficiency is higher for smaller values of second viscosity. The similar result is valid for a regime, when initially only velocity components are perturbed.

It is worth noting that in the current manuscript we have generally examined the role of second viscosity and did not study the self heating mechanism for physical processes implying realistic values of $\eta$ and $\zeta$. This particular problem we are going to consider sooner or later in forthcoming papers.
\begin{figure}[h]
\resizebox{0.5\textwidth}{!}{%
  \includegraphics{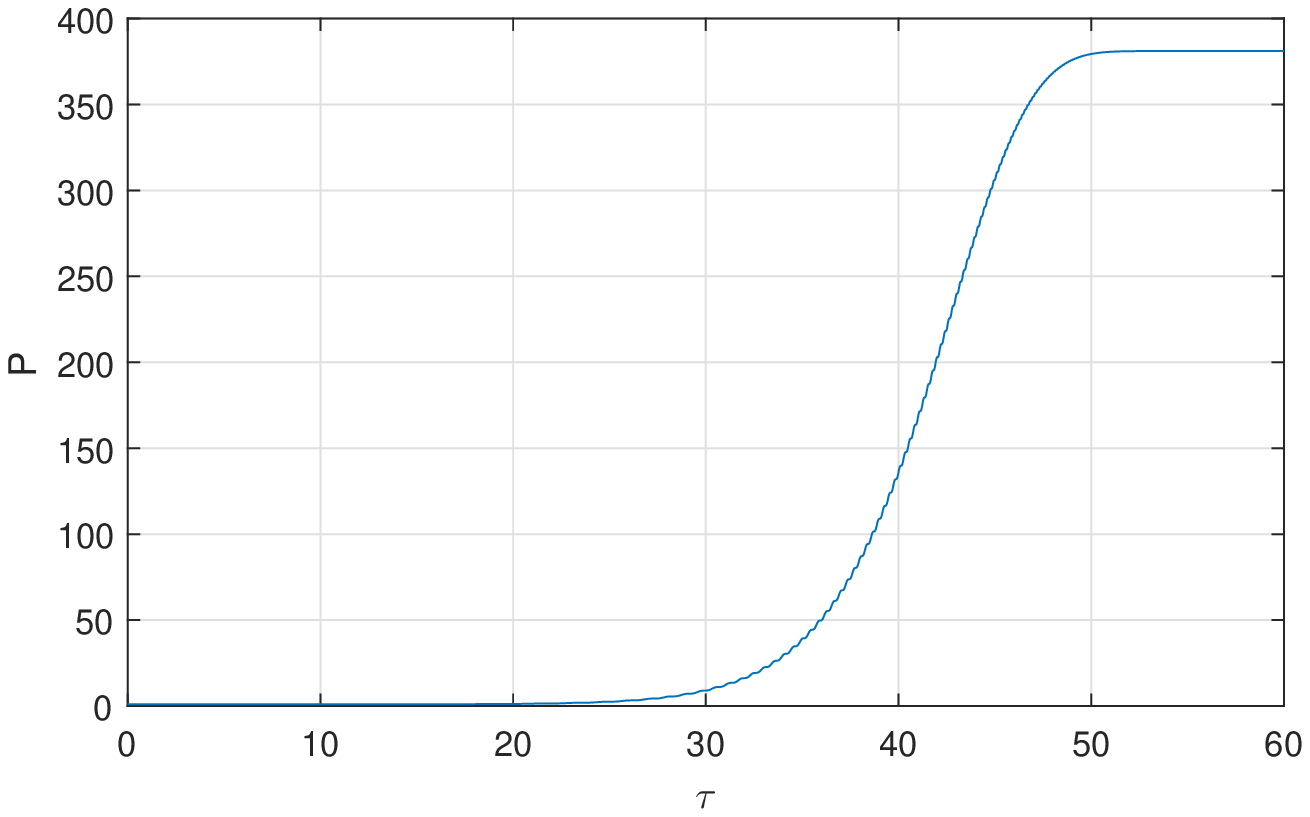}
  \includegraphics{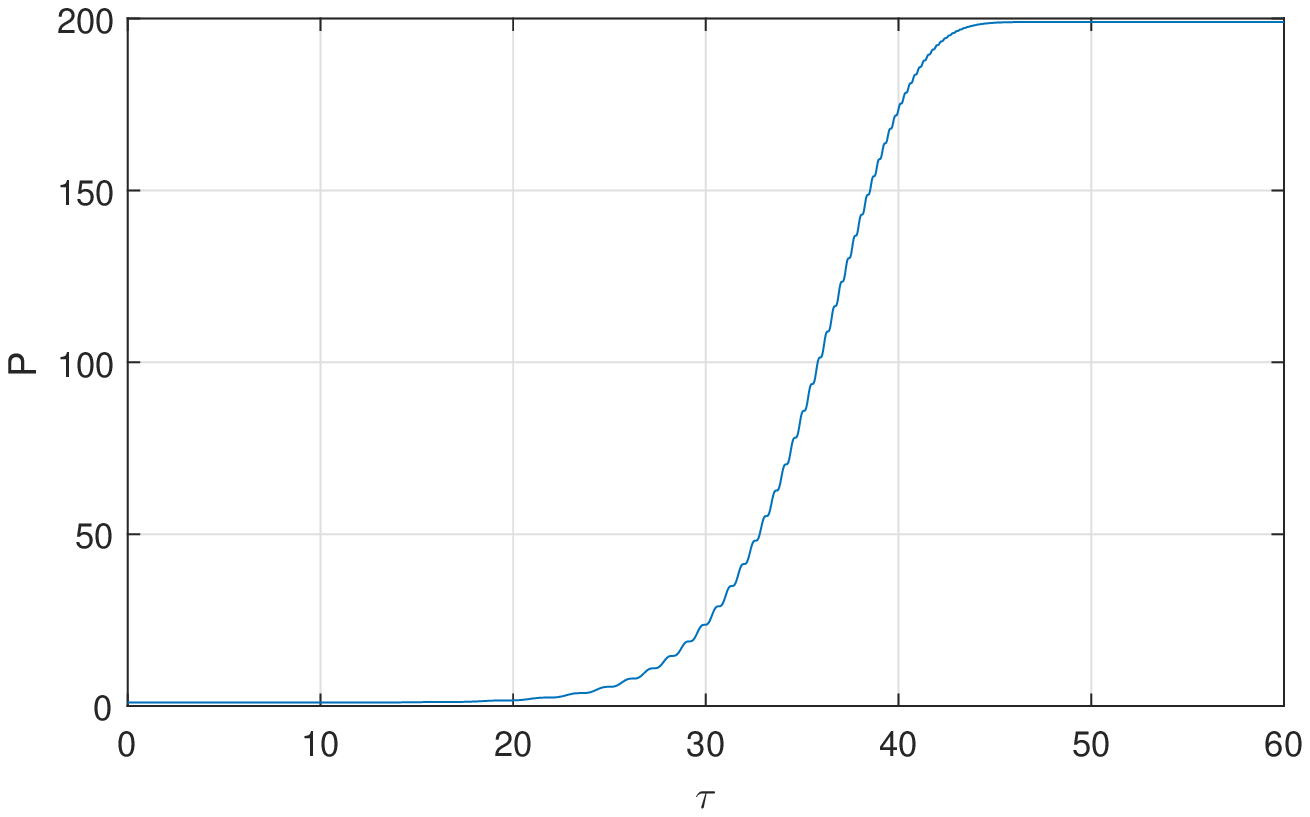}
 }
 \caption{We show the plots of $\psi$ versus time for $\zeta=0$ (left panel) and $\zeta = 0.003$ (right panel) respectively. The rest of the parameters is the same as in Fig. 4.}
 \label{fig5}
\end{figure}

\section{Authors contributions}
All the authors were involved in the preparation of the manuscript.
All the authors have read and approved the final manuscript.

 \section*{Acknowledgments}
The research of GT was supported by the Knowledge Foundation
at the Free University of Tbilisi. The research of ZO was supported by the Shota Rustaveli
National Science Foundation grant (DI-2016-14).

%
% BibTeX users please use
% \bibliographystyle{}
% \bibliography{}

\begin{thebibliography}{}
\bibitem{broder} Broderick, Avery E. \& Loeb, Abraham, ApJ, \textbf{703} (2009) 104L
\bibitem{kharb}
Kharb, P., Gabuzda, D. C., O'Dea, C. P., Shastri, P. \& Baum, S. A., ApJ, \textbf{694} (2009) 1485
\bibitem{yso}
Chrysostomou, A., Bacciotti, F., Nisini, B., Ray, T. P., Eislöffel,
J., Davis, C. J. \& Takami, M., A\&A, \textbf{482} (2008) 575
\bibitem{pm98} Pike C.D., Mason
H.E., Sol. Phys., \textbf{182} (1998) 333
\bibitem{modal} 	
Nold, A. \& Oberlack, M., 2013, PhFl, 25, 104101
\bibitem{tatsuno} 	
Tatsuno, T., Volponi, F. \& Yoshida, Z., 2008, PhPl, 8, 399
\bibitem{kelvin} Lord Kelvin (W. Thomson), Phil. Mag., \textbf{24} (1887)
Ser. 5, 188
\bibitem{tref} Trefethen L.N., Trefethen A.E., Reddy S.C. Driscoll T.A.,
Sience, \textbf{261} (1993) 578
\bibitem{chven} Rogava A.D., Bodo G., Massaglia S. \& Osmanov, Z., A\&A, \textbf{408} (2003)
401
\bibitem{electrost} Osmanov, Z., Rogava A.D., \& Poedts, S., NJP, \textbf{17} (2015)
043019
\bibitem{andro} Rogava A.D., Ap\&SS, \textbf{293} (2004) 189
\bibitem{rop10} Rogava A.D., Osmanov Z. \& Poedts, S., MNRAS,
\textbf{404} (2010) 224
\bibitem{orp12} Osmanov, Z., Rogava, A. D. \& Poedts, S.,
Phys. Plasmas \textbf{19} (2012) 012901
\bibitem{mahrog} Mahajan S. \& Rogava A.D., ApJ, \textbf{518} (1999)
814
\bibitem{landau}
Landau, L. D. \& Lifshitz, E. M., Fluid Mechanics,  Butterworth Heinemann, 2010
\bibitem{cramer} Cramer, M.S., Phys. Fluids, \textbf{24} (2012) 066102

%Author, \textit{Book title} (Publisher, place year) page numbers
% etc
\end{thebibliography}

\end{document}